\documentclass[a4paper]{article}

\usepackage{amssymb}

\newtheorem{algorithm}{Algorithm}[section]
\newtheorem{theorem}{Theorem}[section]
\newtheorem{corollary}{Corollary}[section]
\newtheorem{lemma}{Lemma}[section]

\title{INCLUSION OF REGULAR AND LINEAR LANGUAGES IN GROUP LANGUAGES}

\author{Krasimir Yordzhev}
\date{}

\begin{document}

\pagestyle{myheadings}
\thispagestyle{plain}
\markboth{K. YORDZHEV}{REGULAR, LINEAR AND GROUP LANGUAGES}

\maketitle
\begin{center}
Faculty of Mathematics and Natural Sciences\\
South-West University, Blagoevgrad, Bulgaria\\
E-mail: yordzhev@swu.bg
\end{center}

\begin{abstract}
Let $\Sigma = X\cup X^{-1} = \{ x_1 ,x_2 ,..., x_m ,x_1^{-1} ,x_2^{-1} ,..., x_m^{-1} \}$ and let $G$ be a group with set of generators $\Sigma$. Let $\mathfrak{L} (G) =\left\{ \left. \omega \in \Sigma^* \; \right\vert \;\omega \equiv e \; (\textrm{mod} \; G) \right\} \subseteq \Sigma^*$ be the group language representing $G$, where $\Sigma^*$ is a free monoid over $\Sigma$ and $e$ is the identity in $G$. The problem of determining whether a context-free language is subset of a group language is discussed. Polynomial algorithms are presented for testing whether a regular language, or a linear language is included in a group language. A few finite sets  are built, such that each of them is included in the group language $\mathfrak{L} (G)$ if and only if the respective context-free language is included in $\mathfrak{L} (G)$.
\end{abstract}

\textbf{Key words:} group language; context-free language; regular language; linear language; finite automaton; linear grammar; transition diagram

\textbf{2010 AMS subject classifications:} 68Q45; 68Q70

\section{Introduction}

For more information on automata and language theory we refer the reader to \cite{Hopcroft}. For the mathematical foundations and algebraic approach of formal language theory we refer to \cite{Lallement79,PerrinPin}. For the connections between formal language theory and group theory we recommend the source   \cite{Chiswell}.  List of open problems related to the discussed in this paper topics is given in \cite{Grigorchuk}.

Let
\begin{equation}\label{Sigma}
\Sigma=X \cup X^{-1} = \left\{ x_1 ,x_2,...,x_m ,x_1^{-1} ,x_2^{-1} ,...,x_m^{-1} \right\}
\end{equation}
be an finite  alphabet and let $\Sigma^*$  denote the free monoid over $\Sigma$. Let $G$ be a group with the set of generators $\Sigma$, the set of defining  relations $\Theta$, unit element $e$ and with decidable  word problem. Then the set of words
\begin{equation}\label{L(G)}
\mathfrak{L} (G) =\left\{ \left. \omega \in \Sigma^* \; \right\vert \;\omega \equiv e \; (\textrm{mod} \; G) \right\} \subseteq \Sigma^*
\end{equation}
will be called a group language, which specifies the group G.  The group G is  specified by a context-free language, if the relevant group language  $\mathfrak{L} (G)$ is  context-free. The group G in this case is called a context-free group.

The notion of group language was introduced by A. V. Anisimov in \cite{springerlink:10.1007/BF01071030}. In this article Anisimov proved that $\mathfrak{L} (G)$ is regular if and only if the group $G$ is finite (See also   \cite[Theorem 5.17]{Chiswell}).

A somewhat different definition of the term group language is given in \cite{Heam20115808}, namely a regular language whose syntactic monoid is a finite group. In the given above definition are allowed context-free languages which are not regular. In our work we will stick to the first definition given by A.V. Anisimov.

In \cite{springerlink:10.1007/BF01068773} A.V. Anisimov has showed that the problem of determining the unambiguity of finite automata is a special case of the problem of determining whether a context-free language is a subset of a group language.
Then the question of finding a polynomial algorithms verifying the inclusion of context-free
languages in group languages naturally arises.
This article discusses the most important types of context-free languages - the regular and the linear ones. Regular languages are presented
with the help of finite automata,  and linear languages with the help of linear grammars. In both cases a few finite sets are built, such that each of them is included in the group language $\mathfrak{L} (G)$ if and only if the respective context-free language is included in $\mathfrak{L} (G)$. As a result polynomial algorithms verifying the inclusion of a regular, or a linear language in a group language are presented.

Throughout this article $G$ will be a finitely  generated  group with decidable word problem, and $\Gamma =(N,\Sigma , \Pi , A_1 )$ will  be a context-free grammar that generates the context-free language $L$, ie $L=L(\Gamma )$, where $N$ is the set of variables (nonterminals), $\Sigma = \{x_1 ,x_2,...,x_m ,x_1^{-1} ,x_2^{-1} ,...,x_m^{-1} \}$ is the set of terminals, $\Pi$ the set of productions, and $A_1 \in N$ the start symbol. Let $r$ be the constant of the pumping lemma for context-free language $L$ (see  \cite[Theorem 7.18]{Hopcroft}).

We define the sets:\\
$\displaystyle \Omega_1 =\left\{ \omega\in L \; \left|\; |\omega |\le r \right. \right\} $;\\
$\displaystyle \Omega_2 =\left\{ \left. uwvw^{-1} \; \right\vert \; \vert uwv\vert \le r,\; uv\ne\varepsilon ,\; \exists A\in N : A\stackrel{*}{\Rightarrow} uAv, A\stackrel{*}{\Rightarrow} w  \right\}$;\\
$\displaystyle W_1  =\Omega_1 \cup \Omega_2  $.

The following theorem is proved in \cite{springerlink:10.1007/BF01068773}:

\begin{theorem}\label{anisimov}
{\bf (A. V. Anisimov  \cite{springerlink:10.1007/BF01068773}) }
With the above notation, $L \subseteq \mathfrak{L} (G)$ if and only if  $W_1 =\Omega_1 \cup \Omega_2 \subseteq \mathfrak{L} (G)$.
\hfill $\Box$
\end{theorem}

Theorem \ref{anisimov} gives us an algorithm to check whether the inclusion $L \subseteq \mathfrak{L}(G)$ is true. Unfortunately, this algorithm is not polynomial. The purpose of this work is to show that if $L$ is a regular or a linear language, then Anisimov's algorithm can be transformed so as to be polynomial.

A \emph{transition diagram} is a 4-tuple $H=(V,R,S,l)$, where $(V,R)$ is a directed graph with set of vertices $V$ and multiset of arcs $R\subseteq V\times V =\{ (v_1 ,v_2 )\; |\; v_1 ,v_2 \in V\}$; $S$ is a semigroup whose elements will be called labels and $l$ is a mapping from R to S, which we call labeling mapping. If $\pi=p_1 \; p_2 \; \cdots \; p_k$ is a walk in H, $p_i \in R$, $i=1,2,\ldots k$ then by definition
$$l(p_1 \; p_2 \; \cdots \;p_k )=l(p_1 )l(p_2 )\ldots l(p_k ).$$
If $P$ is a set of walks in $H$, then by definition $$ l(P)=\bigcup_{\pi\in P} l(\pi ) = \{\omega \in S\;\vert \; \exists \pi \in P : l(\pi )=\omega \}$$

Some of the outcomes in this article were announced in Russian in the conference \cite{Nis}.

\section{Inclusion of regular  languages in group languages}

Throughout this section $L$ will mean a regular language.  Then there is a (deterministic or nondeterministic) finite automaton
\begin{equation}\label{A}
A=(Q,\Sigma ,\delta ,q_1 ,Z)
\end{equation}
such that
$$L=L(A)=\{ \omega\in \Sigma^* \; |\; \delta (q_1 ,\omega )\cap Z \ne \emptyset \} ,$$
where:

$Q=\{ q_1 ,q_2 ,\ldots  ,q_{n} \}$ is the  set of states;

$\Sigma = \left\{ x_1 ,x_2,...,x_m ,x_1^{-1} ,x_2^{-1} ,...,x_m^{-1} \right\}$  is the set  of input symbols;

$\delta$ is the transition function;

$q_1 \in Q$ is the start state;

$Z\subseteq Q$ is the set of final (or accepting) states.

Let $H_A =(Q,R,\Sigma^* ,l_A)$ be the transition diagram for the automaton (\ref{A}) (see \cite[p. 48]{Hopcroft}).
Let $G$ be a group with decidable word problem, with the set of generators $\Sigma = \{x_1 ,x_2,...,x_m ,x_1^{-1} ,x_2^{-1} ,...,x_m^{-1} \}$
and unit element $e$. Let
$$H_G = (Q,R,G,l_G )$$
be the transition diagram with the same set of vertices and arcs as in $H_A$, but we consider the labels of arcs as elements of the group $G$.

 We consider the  semiring
 $$F_G =(\mathcal{P}(G) ,\cup , \cdot ,\phi ,\{ e\} ) ,$$
 where $\mathcal{P}(G)$ is the set of subsets of $G$. Operations in $F_G$ are respectively the union and the product of sets, identity is the set $\{ e\}$ that contains only the identity $e$ of $G$, and zero - the empty set $\phi$.

 Let $X,Y\in F_G$. In the semiring $F_G$ we define the next binary operation:
 \begin{equation}\label{[x,y]}
X\star Y =\left\{ xyx^{-1} \; | \; x\in X, y\in Y \right\}
 \end{equation}

We consider the following sets of walks in $H_G$:
\begin{description}
  \item[$P_{ij}$] -- the set of all walks  $\pi\in H_G$ with the initial vertex $q_i \in Q$ and the final vertex  $q_j \in Q$, $1\le i,j\le n$;

   \item[$\widehat{P_{ij}}$] -- the set of all walks $\pi\in H_G$ with the initial vertex $q_i \in Q$, the final vertex  $q_j \in Q$, $1\le i,j\le n$, and in which all vertices are distinct, except possibly $q_i = q_j$. $\widehat{P_{ij}} \subseteq P_{ij}$;

  \item[$P_{iZ}$] -- the set of all walks  $\pi\in H_G$ with the initial vertex $q_i \in Q$ and the final vertex an element of $Z$, $1\le i\le n$;

  \item[$\widehat{P_{iZ}}$] -- the set of all walks  $\pi\in H_G$ with the initial vertex $q_i \in Q$, the final vertex an element of $Z$, $1\le i\le n$, and in which all vertices are distinct (except possibly the initial and final vertices). $\widehat{P_{iZ}} \subseteq P_{iZ}$;

  \item[$O_i $] -- the set of  all walks  $\pi\in H_G$ with the initial vertex and the final vertex $q_i \in Q$, $1\le i\le n$, and in which all vertices are distinct (except initial and final vertices which are $q_i$). $O_i =\widehat{P_{ii}}$.
\end{description}

Obviously $L\subseteq  \mathfrak{L} (G)$ if and only if  $l_G (P_{1Z})=\{ e \}$.

 We consider the sets:\\
$\displaystyle \Omega_3 =  \left\{ l_G (\pi ) \; \left| \; \pi \in \widehat{P_{1Z}} \right. \right\}$;\\
$\displaystyle \Omega_4 =\left\{ uvu^{-1} \; \left| \; \exists q_j \in Q , \pi_1 \in \widehat{P_{1j}} , \pi_2 \in O_j , \pi_3 \in P_{jZ} \; : \; u=l_G (\pi_1 ), v=l_G (\pi_2 ) \right. \right\} $;\\
$\displaystyle W_2 = \Omega_3 \cup \Omega_4  \in F_G $.

We define  the sets of walks $\mathcal{K}_{ij}^k$ in $H_G$, where $i,j\in\{ 1,2,...,n\}$, $k\in\{ 0,1,...,n\}$, $n = \vert Q\vert$ as follows:

\begin{equation}\label{Kij0}
\mathcal{K}_{ij}^0 = \left\{ \rho \; \vert \; \rho =( q_i ,q_j ) \in R \right\}
\end{equation}
\begin{equation}\label{Kijk}
\mathcal{K}_{ij}^k = \mathcal{K}_{ij}^{k-1} \cup \mathcal{K}_{ik}^{k-1} \mathcal{K}_{kj}^{k-1}
\end{equation}

It is easy to see that for all $k\in\{ 0,1,\ldots ,n\}$ none of the walks of $\mathcal{K}_{ij}^k$
passes along an interior vertex $q_s$ where $s > k$.

By definition $\mathcal{K}_{ij}^k$ consists only of walks  with the initial vertex $q_i \in Q$ the final vertex $q_j \in Q$, and may not pass through a vertex $q_s$ where $s\ge k$ or
that passes along a walk $\pi_1$ from $q_i$ to $q_k$, then passes along a walk $\pi_2$ from $q_k$ to $q_j$. None of these walks $\pi_1$ or $\pi_2$
passes along an interior vertex $q_s$ where $s \ge k$.

We consider the following elements of the semiring $F_G$:\\
$\displaystyle \Omega_5 =\left\{l_G (\pi ) \; |\; \pi\in \mathcal{K}_{1t} , q_t \in Z \right\}$;\\
$\displaystyle \Omega_6 =\left\{ l_G (\pi_1 ) \star l_G (\pi_2 ) \; |\; q_j \in Q, q_t \in Z \, :\, \mathcal{K}_{jt}^n \ne \phi ,\pi_1 \in \mathcal{K}_{1j}^n , \pi_2 \in \mathcal{K}_{jj}^n \right\}$, where $''\star ''$ is defined by (\ref{[x,y]}) operation;\\
$\displaystyle W_3 = \Omega_5 \cup \Omega_6  \in F_G $.

It is not difficult to see that
\begin{equation}\label{OOOOmega}
\Omega_3 \subseteq \Omega_5 \quad \textrm{and} \quad \Omega_4 \subseteq \Omega_6 .
\end{equation}

As in $\mathcal{K}_{ij}^k$ is possible existence of a walk containing a cycle or a loop, then in the general case
$\Omega_3 \ne \Omega_5 \quad \textrm{and} \quad \Omega_4 \ne \Omega_6 .$

\begin{theorem}\label{th2}
Let $L$ be a regular language and let $L=L(A)$, where $A$ is defined by (\ref{A})  automaton. Then with the above notation, the following conditions are equivalent:

(i)  $L \subseteq \mathfrak{L} (G)$ ;

(ii)   $W_1 =\Omega_1 \cup \Omega_2 =\{ e\}$ ;

(iii)   $W_2 =\Omega_3 \cup \Omega_4 =\{ e\}$ ;

(iv)   $W_3 =\Omega_5 \cup \Omega_6 =\{ e \}$ .

\end{theorem}

Proof. Since regular languages are special cases of context-free languages, the equivalence of conditions (i) and (ii) was proved by A.V. Anisimov in \cite{springerlink:10.1007/BF01068773} (Theorem \ref{anisimov}). Besides $W_2 \subseteq W_3$ (see  (\ref{OOOOmega})), ie   $W_3 =\{ e\}$ implies  $W_2 =\{ e\}$. So we proved that  (iv) implies  (iii). To prove the theorem we have to prove that (iii) implies  (i) and  (i) implies  (iv).

(iii) implies (i): Let $W_2 =\Omega_3 \cup \Omega_4 =\{ e\}$    and let    $\omega \in L$. Then there is a walk   $\pi \in P_{1Z}$  such that  $l_A (\pi )=\omega$.

If $\pi$ does not contain cycles and loops, then $l_G (\pi ) \in l_G (\widehat{P_{1Z}} ) =\Omega_3 =\{ e\}$  and therefore $\omega \in \mathfrak{L} (G)$.

Let $\pi$ contains a cycle or a loop. In other words, there is $q_j \in Q$ such that $\pi$ can be expressed as $\pi =\pi_1 \pi_2 \pi_3$, where $\pi_1 \in \widehat{P_{1j}} , \pi_2 \in O_j , \pi_3 \in P_{jZ}$ and $l_G (\pi_1 )l_G (\pi_2 )(l_G (\pi_1 ))^{-1} \in \Omega_4 =\{ e\}$. Therefore, $l_G (\pi_1 )l_G (\pi_2 )=l_G (\pi_1 )$ and $l_G (\pi_1 \pi_2 \pi_3 )=l_G (\pi_1 \pi_3 )$. Since $\pi_2 \in O_j$, then the length of $\pi_2$ is greater than 1. Consequently, in $H_G$ there is a walk with less length than the length of $\pi$, whose label is equal to $\omega$ in the group $G$. This process of reduction may proceed a finite number of times as the length of $\omega$ is finite. At the end of this process we obtain a walk in $H_G$ without cycles and without loops with label equal to $\omega$ as an element of the group $G$. But $l_G (\widehat{P_{1Z}} )=\Omega_3 =\{ e\}$. Hence $\omega =e$ in the group $G$ and therefore $L \subseteq \mathfrak{L} (G)$.

(i) implies  (iv): Let $L \subseteq \mathfrak{L} (G)$ ie $l_G (P_{1Z} )=\{ e\}$. From $\Omega_5 \subseteq l_G (P_{1Z})$ follows  $\Omega_5 =\{ e\}$.

Let $z\in \Omega_6$. Then $z$ can be represented in the form $z=uvu^{-1}$, where $u\in l_G (\mathcal{K}_{1j}^n) ,$ $v\in l_G (\mathcal{K}_{jj}^n )$ for some integer $j$ such that there is a walk $\pi_3 \in P_{jZ}$ and let $l_G (\pi_3 )=w$. Obviously there are a walk $\pi_1 \in P_{1j}$ and a walk $\pi_2 \in O_j$ such that $u=l_g (\pi_1 )$ and $v=l_g (\pi_2 )$. Thus $\pi ' =\pi_1 \pi_2 \pi_3 \in P_{1Z}$    and $\pi '' =\pi_1 \pi_3 \in P_{1Z}$. Since $L \subseteq \mathfrak{L} (G)$ then $l_G (\pi ')=l_G (\pi '')=e$, therefore $uvw=uw$, ie  $ uvu^{-1} =e$. Hence $z=e$ and since $z\in \Omega_6$ is arbitrary, then $\Omega_6 =\{ e\}$. The theorem is proved.
\hfill $\Box$

The following algorithm is based on the equivalence (i) and (iv) of Theorem \ref{th2}.  For convenience, $i\in Z$ will mean $q_i \in Z$, and $g_{ij}^k$ will be $l_G (\mathcal{K}_{ij}^k )$. Here, $k$ in $g_{ij}^k$ is a superscript and does not mean an exponent.

\begin{algorithm}\label{alg1}
Verifies the inclusion $L \subseteq \mathfrak{L} (G)$ for a regular language $L$, and a group language $\mathfrak{L} (G)$, where $G$ is a group with  decidable word problem.
\end{algorithm}

{\bf Input:} $g_{ij}^0 =l_G (\mathcal{K}_{ij}^0 ),\quad i,j=1,2,...,n$

{\bf Output:} Boolean variable $T$, which receives the value {\bf True} if $L \subseteq \mathfrak{L} (G)$, and the value {\bf False}, otherwise. The algorithm will stop immediately after the value of $T: = \textbf{False}$.

Begin

1. $T:= \textbf{True} $;

2. For  $1\le k\le n$       Do

3. \hspace{0.4cm}  For $1\le i,j\le n$  Do

4. \hspace{0.8cm} $g_{ij}^k :=g_{ij}^{k-1} \cup g_{ik}^{k-1} g_{kj}^{k-1}$;

5. \hspace{2.0cm}  End Do;

6. \hspace{1.6cm}  End Do;

7. For  $j\in Z$  Do

8. If $g_{1j}^n \ne \phi$  and $g_{1j}^n \ne \{ e\}$   Then

9. \hspace{2.0cm}  Begin  $T:=\textbf{False}$; Halt; End;

10.\hspace{1.2cm}  End Do;

11. For $1\le j\le n$   Do

12.\hspace{0.4cm} For $t\in Z$   Do

13.\hspace{0.4cm} If $g_{jt}^n \ne \phi$ and $g_{1j}^n \ne \phi$ and $g_{jj}^n \ne \phi$  Then

14.\hspace{0.8cm} If $g_{1j}^n \star g_{jj}^n \ne \{ e\}$  Then

15.\hspace{2.0cm}  Begin  $T:=\textbf{False}$; Halt; End;

16.\hspace{1.6cm}  End Do;

17.\hspace{1.2cm}  End Do;

End.

\begin{theorem}\label{tttt3}
Algorithm \ref{alg1} checks the inclusion $L\subseteq \mathfrak{L} (G)$, where $L$ is a regular language recognized by a finite automaton with $n$ states, $\mathfrak{L} (G)$ is a group language, which specifies the group $G$ with  decidable word problem.
Algorithm \ref{alg1} executes at most  $O(n^3 )$ operations $ \cup $ and $\cdot$, and at most $O(n^2 )$ operations $ \star $ in the semiring $F_G$, where the binary operation  $ \star $ is defined using the formula (\ref{[x,y]}).
\end{theorem}

Proof.  According to Theorem \ref{th2} and considering axioms of the semiring  $F_G$, then in rows 9 and 15 of Algorithm \ref{alg1}, the boolean variable $T$ gets the value \textbf{False} if and only if $L$ is not included in $\mathfrak{L} (G)$. Otherwise, $T$ gets the value \textbf{True}. Hence the algorithm correctly checks whether the inclusion $L\subseteq \mathfrak{L} (G)$ is true.

 It is easy to see that line 4 is executed no more than $n^3$ times. The operations $\cup $ and $ \cdot $ (once each of them) in the semiring  $F_G$ is performed during each iteration. Lines 13 and 14 is executed at most $n^2$ times each. Therefore, Algorithm \ref{alg1} performs no more than $O(n^3 )$ operations  $ \cup $ and $ \cdot $, and no more than $O(n^2 )$ operations $\star $  in the semiring $F_G$. The theorem is proved.
\hfill $\Box$

\begin{corollary}
If the operations  $\cup $, $\cdot $ and $\star $ in the semiring $F_G$  can be done in a polynomial time, then Algorithm \ref{alg1} is polynomial.
\end{corollary}

\section{Inclusion of linear languages in group languages}
Let S be an arbitrary monoid with identity 1. We consider the set
$$U_S =S\times S=\{ (x,y) \vert x,y\in S\}.$$

We introduce the operation $\diamond $ in  $U_S$ as follows:  if $(x,y)$, $(z,t) \in U_S$ then
\begin{equation}\label{diamond}
(x,y)\diamond (z,t) = (xz,ty).
\end{equation}

It is easy to see that the operation $\diamond $ is associative and $U_S$  with this operation is a monoid with identity (1,1). If $S$ is a group, then $U_S$ is a group, and if  $a=(x,y)\in U_S$ then the inverse element of $a$ will be  $a^{-1} =(x^{-1} ,y^{-1} )$. We define mappings $f_l $, $f_r $ and $f_d$   from $U_S$  to $S$  as follows:
\begin{equation}\label{fl}
f_l (x,y)=x
\end{equation}
\begin{equation}\label{fr}
f_r (x,y)=y
\end{equation}
\begin{equation}\label{fd}
f_d (x,y)=xy
\end{equation}

Obviously $$f_d (x,y)=f_l (x,y)f_r (x,y) .$$

In this section we consider a linear grammar
\begin{equation}\label{Gamma_lilear}
\Gamma =(N,\Sigma , \Pi , A_1 ) ,
\end{equation}
where:

$N=\{ A_1 ,A_2 ,...,A_n \}$ is the  set of variables (nonterminals);

$\Sigma = \left\{ x_1 ,x_2,...,x_m ,x_1^{-1} ,x_2^{-1} ,...,x_m^{-1} \right\}$  is the set  of input symbols;

$\Pi$ is the set of productions;

$A_1 \in N$ is the start variable.

A context-free grammar $\Gamma =(N,\Sigma , \Pi , A_1 )$ is called \emph{linear} if all productions in $\Pi$ are of the form $A_i \to \alpha A_j \beta$ or $A_i\to \alpha$, where $A_i ,A_j \in N$, $1\le i,j\le n$, $\alpha ,\beta \in \Sigma^*$. A language $L$ is called \emph{linear} if there is a linear grammar $\Gamma$ such that $L=L(\Gamma ,A_1)$.

We consider the transition diagram
\begin{equation}\label{H_Gamma}
H_{\Gamma} =(V,R,U_{\Sigma^*} ,l_{\Gamma} )
\end{equation}
with the set of vertices $V=N\cup \{ A_{n+1} \}$, where $A_{n+1} \notin N$. $U_{\Sigma^*}$ is the considered above monoid with the set of elements $\{ (\alpha ,\beta ) \vert \alpha ,\beta \in \Sigma^* \}$ and with the operation $\diamond $. The set of arcs  $R$ in $H_\Gamma$ is formed as follows:

a) if a  production $A_i \to \alpha A_j \beta$  exists in $\Pi$ where $A_i ,A_j \in N$,  then there exists an arc from $A_i$ to $A_j$   labeled $(\alpha ,\beta )$;

b) if a  production $A_i \to \alpha $  exists in $\Pi$ where $A_i \in N$, $\alpha \in \Sigma^*$,  then there exists an arc from $A_i$ to $A_{n+1}$   labeled $(\alpha ,\varepsilon )$, $\varepsilon$ is the empty word;

c) there are no other arcs in $R$.

Let $G$ be a group with the set of generators  $\Sigma  = \{ x_1 ,x_2 ,...,x_m ,x_1^{-1} ,...,x_m^{-1} \} ,$ with the set of defining  relations $\Theta$, identity $e$ and with decidable  word problem. Let $U_G$ be the group obtained as described above. We consider the transition diagram
\begin{equation}\label{H_U}
H_U =(V,R,U_G ,l_U ),
\end{equation}
 where the set of vertices $V$ and the set of arcs $R$ coincide with the corresponding sets in the transition diagram $H_{\Gamma}$   according to (\ref{H_Gamma}), and labels will be elements of the group $U_G$.

We consider the following sets of walks in $H_U$:
\begin{description}
  \item[$D_{ij}$] -- the set of all walks  $\pi\in H_U$ with the initial vertex $A_i \in V$ and the final vertex  $A_j \in V$, $1\le i\le n$, $1\le j\le n+1$;

   \item[$\widehat{D_{ij}}$] -- the set of all walks $\pi\in H_U$ with the initial vertex $A_i \in V$, the final vertex  $A_j \in V$, $1\le i\le n$, $1\le j\le n+1$, and in which all vertices are distinct, except possibly $A_i = A_j$. $\widehat{D_{ij}} \subseteq D_{ij}$;

  \item[$C_i$] -- the set of  all walks  $\pi\in H_U$ with the initial vertex and the final vertex $A_i \in V$, $1\le i\le n$, and in which all vertices are distinct (except initial and final vertices which are $A_i$). $C_i =\widehat{D_{ii}}$.
\end{description}

\begin{lemma}\label{lll1}
Let $\Gamma =(N,\Sigma , \Pi , A_1 )$ be a linear grammar and $H_\Gamma$ be the transition diagram  according to (\ref{H_Gamma}). Let $P_\Gamma$ be the set of all walks  $\pi\in H_\Gamma$ with the initial vertex $A_1$ and the final vertex  $A_{n+1}$ Then
$$L=L(\Gamma )=f_d (l_{\Gamma} (P_\Gamma )) .$$
\end{lemma}

Proof. Immediate.
\hfill $\Box$

\begin{corollary}\label{ccccc2}
Let $L$ be a  linear language generated by the linear grammar (\ref{Gamma_lilear})  and let $G$ be a finitely  generated  group  with decidable word problem and with the set of generators  $\Sigma  = \{ x_1 ,x_2 ,...,x_m ,x_1^{-1} ,...,x_m^{-1} \}$. Then
$$L\subseteq \mathfrak{L} (G) \Longleftrightarrow  f_d (l_U (D_{1\> n+1} ))=\{ e\}.$$

\end{corollary}

As in section 2, we can consider the semirings  $F_G =(\mathcal{P}(G) ,\cup , \cdot ,\phi ,\{ e\} )$ and  $F_U =(\mathcal{P}(U_G ),\cup ,\diamond ,\phi ,\{ (e,e) \} )$. Defined using equations (\ref{fl}), (\ref{fr}) and (\ref{fd}) mappings $f_l$, $fr$, and $f_d$ can be extended in a natural way to mappings from $F_U$ to $F_G$.

Let $X,Y,Z\in F_G$. In $F_G$, we introduce the next operation:
\begin{equation}\label{xxyyzz}
\langle X,Y,Z\rangle =\left\{ xyzy^{-1} \; |\; x\in X,y\in Y,z\in Z\right\}
\end{equation}

We consider the following elements of the semiring $F_G$:\\
$\displaystyle \Omega_7 =\left\{ f_d (l_U (\pi )) \; \left| \; \pi\in\widehat{D_{1\, n+1}} \right.\right\} $;\\
$\displaystyle \Omega_8 =\displaystyle \left\{ \langle f_l (l_U (\pi_2 )),f_d (l_U (\pi_3 )),f_r (l_U (\pi_2))\rangle \; \left\vert \; \exists \pi\in D_{1\, n+1} \; :\;  \pi =\pi_1 \pi_2 \pi_3  ,\right.\right. $

    $\displaystyle \left.  \pi_1 \in D_{1i} ,\, \pi_2 \in C_i ,\, \pi_3 \in \widehat{D_{i\, n+1}} ,\, 1\le i\le n \right\} $;\\
$\displaystyle W_4 =\Omega_7 \cup \Omega_8 $.

It is not difficult to see that
\begin{equation}\label{W4W1}
\Omega_7 \subseteq \Omega_1 \quad \textrm{and} \quad \Omega_8 \subseteq \Omega_2 \Longrightarrow W_4 \subseteq W_1
\end{equation}
and in the general case
$\Omega_7 \ne \Omega_1 \quad \textrm{and} \quad \Omega_8 \ne \Omega_2 .$

As in section 2 (see (\ref{Kij0}) and (\ref{Kijk})), we define  the sets of walks $\mathcal{K}_{ij}^k$ in $H_U$, where $i\in\{ 1,2,...,n\}$, $j\in\{ 1,2,...,n+1\}$,  $k\in\{ 0,1,...,n\}$, $n = \vert N \vert$, $V=N\cup \{ A_{n+1} \}$,  $N=\{ A_1 ,A_2 ,...,A_n \}$ is the  set of variables of the grammar $\Gamma$, $A_{n+1} \notin N$.

\begin{equation}\label{KUij0}
\mathcal{K}_{ij}^0 = \left\{ \rho \; \vert \; \rho =( A_i ,A_j ) \in R \right\}
\end{equation}
\begin{equation}\label{KUijk}
\mathcal{K}_{ij}^k = \mathcal{K}_{ij}^{k-1} \cup \mathcal{K}_{ik}^{k-1} \mathcal{K}_{kj}^{k-1}
\end{equation}

Let $g_{ij}^k =l_U (\mathcal{K}_{ij}^k ) \in F_U$, where  $k$ is a superscript and does not mean an exponent.

 We consider the following elements of the semiring $F_G$:\\
$\displaystyle \Omega_9 = f_d (g_{1\, n+1}^n ) $;\\
$\displaystyle \Omega_{10}$] $=\left\{ \langle f_l (g_{ii}^n ) , f_d (g_{i\ n+1}^n ) , f_r (g_{ii}^n )\rangle \; \left| \; 1\le i\le n,\, \mathcal{K}_{1i}^n \ne \emptyset ,\mathcal{K}_{ii}^n \ne \emptyset ,\mathcal{K}_{i\, n+1}^n \ne \emptyset \right.\right\} $;\\
$\displaystyle W_5 =\Omega_9 \cup \Omega_{10} $.

It is easy to see that
\begin{equation}\label{W4W5}
\Omega_7 \subseteq \Omega_9 \quad \textrm{and} \quad \Omega_8 \subseteq \Omega_{10} \Longrightarrow W_4 \subseteq W_5.
\end{equation}

As in $\mathcal{K}_{ij}^k$ is possible existence of a walk containing a cycle or a loop, then in the general
$\Omega_7 \ne \Omega_9 \quad \textrm{and} \quad \Omega_8 \ne \Omega_{10} .$

\begin{theorem}\label{th3}
Let $L$ be a linear language and let $L=L( \Gamma )$, where $\Gamma$ is defined by (\ref{Gamma_lilear}) linear grammar. Then with the above notation, the following conditions are equivalent:

(i)  $L\subseteq \mathfrak{L} (G)$ ;

(ii) $W_1 =\Omega_1 \cup \Omega_2 =\{ e \}$ ;

(iii) $W_4 =\Omega_7 \cup \Omega_8 =\{ e \}$ ;

(iv)  $W_5 =\Omega_9 \cup \Omega_{10} =\{ e \}$ .
\end{theorem}

Proof. The equivalence of conditions (i) and (ii) was proved by A.V. Anisimov in \cite{springerlink:10.1007/BF01068773} (Theorem \ref{anisimov}). As noted above (see \ref{W4W1}),  $W_4 \subseteq W_1$ and hence $W_1 =\{ e \}$ implies $W_4 =\{ e\}$, ie (ii) implies (iii).  From (\ref{W4W5}) follows that (iv) implies (iii). To prove the theorem it is sufficient to prove that (iii) implies  (i) and  (i) implies  (iv).

(iii) implies (i): Let  $W_4 =\Omega_7 \cup \Omega_8 = \{ e\}$   and let  $\omega \in L$. According to Lemma \ref{lll1} $\omega \in f_d (l_{\Gamma} (P_{\Gamma} ))$. Hence $\omega$ can be written as $\omega =\omega_1 \omega_2$, where  $(\omega_1 ,\omega_2 )$ is the label of a walk in $H_\Gamma$ with the initial vertex $A_1$ and the final vertex  $A_{n+1}$ and let $\pi$ be the corresponding path in $H_U$.  $f_d (l_U (\pi  )) \equiv \omega$ (mod $G$)  is satisfied.

If  $\pi\in\widehat{D_{1\, n+1}}$   then     $f_d (l_U (\pi )) \in f_d (l_U (\widehat{D_{1\, n+1}} )) = \Omega_7 \subseteq \{ e\}$ and therefore    $\omega \in \mathfrak{L} (G)$.

Suppose $\pi$ contains a cycle or a loop. Then $\pi$ can be written as $\pi = \pi_1 \pi_2 \pi_3$, where $\pi_1 \in D_{1j}$, $\pi_2 \in C_j$ and $\pi_3 \in \widehat{D_{j\, n+1}}$ for some $j\in \{ 1,2,\ldots ,n\}$. Let    $l_U (\pi_1 ) = (a_1 ,b_1 )$, $l_U (\pi_2 ) = (a_2 ,b_2 )$ and $l_U (\pi_3 ) = (a_3 ,b_3 )$. Then
$$f_d (l_U (\pi )) = f_d ((a_1 ,b_1 )\diamond (a_2 ,b_2 )\diamond (a_3 ,b_3 )) = f_d (a_1 a_2 a_3 ,b_3 b_2 b_1) = a_1 a_2 a_3 b_3 b_2 b_1 .$$
But $a_2 a_3 b_3 b_2 (a_3 b_3 )^{-1} \in \Omega_8$,  hence  $a_2 a_3 b_3 b_2 (a_3 b_3 )^{-1} = e$, ie  $a_1 a_2 a_3 b_3 b_2 b_1 = a_1 a_3 b_3 b_1$. It is easy to see that $(a_1 a_3 ,b_3 b_1 )$ is the label of the walk $\pi_1 \pi_3$, which is obtained from $\pi$ by omitting  $\pi_2$. We continue to omit the cycles and loops in $\pi$. Because the word $\omega$ is finite, after finitely many steps we obtain a walk $\pi ' \in \widehat{D_{1\, n+1}}$ such that  $f_d (l_u (\pi ')) = f_d (l_U (\pi  )) \equiv \omega$ (mod $G$).  But $f_d (l_U (\pi ' ))\in \Omega_7 = \{ e\}$. Therefore $f_d (l_U (D_{1\> n+1} ))=\{ e\}$ and according to Corollary \ref{ccccc2} $L\subseteq \mathfrak{L} (G)$.

(i) implies (iv): Let  $L\subseteq \cal M$. Then according to Corollary   \ref{ccccc2}   $f_d (l_U (D_{1\, n+1} ))=\{ e\}$. It is obvious that $\mathcal{K}_{1\ n+1}^n \subseteq D_{1\, n+1}$   and therefore    $\Omega_9 =\{ e\}$.

Let $z\in \Omega_{10}$. Then   $z=uvwv^{-1}$, where $u\in f_l (g_{ii}^n )$, $v=f_d (g_{i\ n+1}^n )$, $w=f_r (g_{ii}^n )$  and  there are walks $\pi_1 \in D_{1i}$, $\pi_2 \in C_i$, $\pi_3 \in D_{i\, n+1}$ for some  $i\in \{ 1,2,\ldots ,n \}$  such that $(u,w)=l_u (\pi_2 )$ and  $v=v_1 v_2$, where $(v_1 ,v_2) = l_u (\pi_3 )$. Let $l_U (\pi_1 )=(x,y)$.  We consider the walks $\pi ' =\pi_1 \pi_2 \pi_3 \in D_{1\, n+1}$ and  $\pi '' =\pi_1 \pi_3 \in D_{1\, n+1}$. We have:

$l_U (\pi ')=l_U (\pi_1 \pi_2 \pi_3 ) = (x,y)\diamond (u,w) \diamond (v_1 ,v_2 ) = (xuv_1 ,v_2 wy)$

$l_U (\pi '')=l_U (\pi_1 \pi_3 ) = (x,y)\diamond (v_1 ,v_2 ) = (xv_1 ,v_2 y)$

According to Corollary 3 $xuvwy=xvy=e$, which implies that  $uvwv^{-1} =e$, ie $z=e$. Since $z\in \Omega_{10}$ is an arbitrary $z$, then  $\Omega_{10} =\{ e\}$. The theorem is proved.

\hfill $\Box$

The following algorithm is based on the equivalence (i) and (iv) of Theorem \ref{th3}.

\begin{algorithm}\label{alg2}
Verifies the inclusion $L \subseteq \mathfrak{L} (G)$ for a linear language $L$, and a group language $\mathfrak{L} (G)$, where $G$ is a group with  decidable word problem.
\end{algorithm}

{\bf Input:} $g_{ij}^0 =l_U (\mathcal{K}_{ij}^0 )$, $i=1,2,...,n$, $j=1,2,...,n,n+1$

{\bf Output:} Boolean variable $T$, which receives the value {\bf True} if $L \subseteq \mathfrak{L} (G)$, and the value {\bf False}, otherwise. The algorithm will stop immediately after the value of $T: = \textbf{False}$.

begin

1. $T:=$ \textbf{True} ;

2. For $1\le k \le n$      Do

3. \hspace{0.4cm}    For $1\le i\le n$ and $1\le j\le n+1$   Do

4. \hspace{0.8cm} $g_{ij}^k :=g_{ij}^{k-1} \cup g_{ik}^{k-1} \diamond g_{kj}^{k-1}$;

5. \hspace{2.0cm}       End Do;

6. \hspace{1.6cm}       End Do;

7. If $g_{1\ n+1}^n \ne \phi$  и $f_d (g_{1\ n+1}^n )\ne \{ e\}$     Then

8. \hspace{1.2cm}     Begin $T:=$ \textbf{False};  Halt;    End;

9. For $1\le i\le n$      Do

10. \hspace{0.4cm}    If $g_{1i}^n \ne \phi$  and $g_{ii}^n \ne \phi$   and  $g_{i\ n+1}^n \ne \phi$     Then

11. \hspace{0.8cm}   If $\langle f_l (g_{ii}^n ),f_d (g_{i\ n+1}^n ),f_r (g_{ii}^n )\rangle \ne \{ e \}$   Then

12. \hspace{1.2cm}   Begin $T:=$ \textbf{False}; Halt;    End;

13.\hspace{1.2cm}     End Do;

End.

\begin{theorem}\label{tttt5}
Algorithm \ref{alg2} checks the inclusion $L\subseteq \mathfrak{L} (G)$, where $L$ is a linear language generated by a linear grammar  with $n$ variables, $\mathfrak{L} (G)$ is a group language, which specifies the group $G$ with  decidable word problem.
Algorithm \ref{alg2} executes at most  $O(n^3 )$ operations $ \cup $ and $\diamond$ in the semiring $F_U$, and no more than $O(n^2 )$ operations $\langle \ ,\ ,\ \rangle$ in the semiring $F_G$, where the operation  $\langle \ ,\ ,\ \rangle$ is defined using the formula (\ref{xxyyzz}).
\end{theorem}

Proof. Similarly as the proof of Theorem \ref{tttt3}.
\hfill $\Box$

\begin{corollary}
If the operations  $\cup $, $\diamond$ in the semiring $F_U$ and the operation $\langle \ ,\ ,\ \rangle$  in the semiring $F_G$  can be done in a polynomial time, then Algorithm \ref{alg1} is polynomial.
\end{corollary}

\end{document}